\documentclass[aps,pra,twocolumn,showpacs,superscriptaddress, nofootinbib]{revtex4-2}

\usepackage[english]{babel}
\usepackage[utf8]{inputenc}

\usepackage[pdftex, pdftitle={Article}, pdfauthor={Author}]{hyperref} 

\usepackage{natbib}
\usepackage{hyperref}
\usepackage{verbatim}
\usepackage{amsmath}
\usepackage{amsfonts}
\usepackage{amssymb}
\usepackage{amsthm}
\usepackage{mathtools}
\usepackage{bm}
\usepackage{xparse}
\usepackage{graphicx}
\usepackage{xcolor}
\usepackage{textcase}
\usepackage{url}
\usepackage{lipsum}
\usepackage{appendix}
\usepackage{epstopdf}
\usepackage{lipsum} 
\usepackage[]{mdframed}
\usepackage{dsfont}
\def\ket#1{|#1\rangle}
\def\bra#1{\langle#1|}

\def\Tr{\mathrm{Tr}}

\newcommand{\diagdots}[3][-25]{%
  \rotatebox{#1}{\makebox[0pt]{\makebox[#2]{\xleaders\hbox{$\cdot$\hskip#3}\hfill\kern0pt}}}%
}

\DeclareDocumentCommand{\Tr}{m m O{\big}}{{\rm Tr}_{\:\!{#1}}#3({#2}#3)}

\begin{document}
\title{Wigner's friend scenarios: on what to condition and how to verify the predictions}

\author{Flavio Del Santo}
\affiliation{Vienna Center for Quantum Science and Technology (VCQ), Faculty of Physics, Boltzmanngasse 5, University of Vienna, Vienna A-1090, Austria.}
\affiliation{Group of Applied Physics, University of Geneva, 1211 Geneva, Switzerland}
\affiliation{Institute for Quantum Optics and Quantum Information (IQOQI),
Austrian Academy of Sciences, Boltzmanngasse 3, A-1090 Vienna, Austria.}

\author{Gonzalo Manzano}
\affiliation{Institute for Cross-Disciplinary Physics and Complex Systems, (IFISC, UIB-CSIC), Campus Universitat de les Illes Balears E-07122, Palma de Mallorca, Spain} 

\author{\v Caslav Brukner}
\affiliation{Vienna Center for Quantum Science and Technology (VCQ), Faculty of Physics, Boltzmanngasse 5, University of Vienna, Vienna A-1090, Austria.}
\affiliation{Institute for Quantum Optics and Quantum Information (IQOQI),
Austrian Academy of Sciences, Boltzmanngasse 3, A-1090 Vienna, Austria.}

\date{\today}

\begin{abstract}

Wigner's friend experiment and its modern extensions display the ambiguity of the quantum mechanical description regarding the assignment of quantum states.  While the friend applies the state-update rule to the system upon observing an outcome of her measurement in a quantum system, Wigner describes the friend's measurement as a unitary evolution, resulting in an entangled state for the composite system of the friend and the system. In this respect, Wigner is often referred to as a “superobserver" who has the supreme technological ability to keep the friend's laboratory coherent. As such, it is often argued that he has the “correct" description of the state. Here we show that the situation is more symmetrical than is usually thought: there are different types of information that each of the observers has that the other fundamentally cannot have - they reside in different “bubbles" (in Calvalcanti's terminology). While this can explain why the objectivity of the state assignment is only relative to the bubble, we consider more elaborated situations in the form of a game in which the players can switch between bubbles. We find that, in certain circumstances, observers may be entitled to adopt and verify the state assignment from another bubble if they condition their predictions on \textit{all} information that is in principle available to them.


\end{abstract}
\maketitle

\section{Introduction}

The notorious \textit{quantum measurement problem} stems from an ambiguity between two alternative dynamics that are present in quantum theory. In the words of Baumann and Wolf
~\cite{Baumann2018}, “in what is called \textit{standard quantum mechanics}, the measurement-update rule, commonly associated with a collapse, is a break with the otherwise unitary evolution governed by the Schr\" odinger equation. The formalism, however, provides no indication about when to apply this rule: It does not state what qualifies some interactions as measurements but not others (the measurement problem)."

The problem is most notably exemplified by the \textit{Wigner's friend} thought experiment \cite{Wigner1995}, whose scope has been enlarged in recent years \cite{Brukner2015, Brukner2018, Frauchiger2018, healey2018quantum, baumann2019comment, baumann2020wigner, Baumann2019, proietti2019experimental, bong2020strong,  cavalcanti2021view, zukowski2021physics, guerin2021no, leegwater2022greenberger, ormrod2022no, vilasini2022general, haddara2023possibilistic,  wiseman2023thoughtful,   baumann2023observers, baumann2023classical, utreras2024allowing, polychronakos2024quantum}. This takes the ambiguity about which dynamical law to apply to the extreme by featuring an observer---historically referred to as the \textit{friend}---who takes both the roles of an observer performing a quantum experiment and of a quantum system observed by another observer, \textit{Wigner}. More specifically, the friend performs a measurement on a quantum system in her laboratory, while Wigner is able to perform quantum measurement on the whole laboratory, thus treating both the friend and the particle as a joint quantum system.

What is the state of the system after the friend's measurement? Since the friend observes a definite outcome, it seems legit to say that the state should be updated accordingly, thus applying the “collapse" rule. On the other hand, from Wigner's perspective, the friend and the system are merely two interacting subsystems of an isolated system, thus he seems entitled to describe the same physical process unitarily. The question then arises: is there a single, absolute, “correct" state? 
By this, we mean that \textit{every} hypothetical observer should assign the same state, which can then be used to make empirically verifiable predictions consistent with quantum theory.

The answer in the positive view has been recently challenged by a series of no-go theorems \cite{Brukner2015, Frauchiger2018, Brukner2018,  bong2020strong, cavalcanti2021view, wiseman2023thoughtful, leegwater2022greenberger, haddara2023possibilistic, baumann2023observers}, in so far as it is possible that there exist no “single reality" that can be shared between the friend and Wigner. It was proposed instead that Wigner and his friend could live in different, incommensurable realities, called “bubbles" \cite{cavalcanti2021view}.\footnote{The concept of bubble, as it will be used here, is very similar to Healey's “agent situation" \cite{healey2012quantum}. These ideas are reminiscent of the philosophical position of \emph{fragmentalism} \cite{fine2006reality,  lipman2015fine}, which upholds “that the world is not a monolithic whole constructed from mutually compatible facts, but rather a collection of \textit{fragments}, with each fragment containing mutually compatible facts, while different fragments are incompatible" \cite{dieks2022perspectival}.}

In this work, we (i) refine and further formalize the concept of a “bubble” as an information-theoretic notion, and model communication between bubbles; (ii) devise a scenario (in the form of a gambling game) in which an observer living in one bubble can be persuaded that their state assignment---based solely on the information available from measurement results obtained within their own bubble---is not optimal for predicting future measurements, whereas the state assignment of an observer in another bubble is; and (iii) demonstrate, for the first time, that the friend---typically regarded as less powerful than the superobserver Wigner---can make more precise predictions than the latter, thereby symmetrizing the narrative. 

\section{“Information Bubbles"}
\label{bubbles}
To describe the incommesurability of the measurement outcomes by different observers in Wigner's friend extended scenarios, Cavalcanti introduced the concept of “bubbles" \cite{cavalcanti2021view}.
Building on this intuition, a bubble can be defined as the "locus" where the same relevant information---for example, a measurement result---is, in principle, available.
Every hypothetical observer who has in principle access to the same relevant classical information---i.e., that can be copied, broadcasted, etc.---lives, therefore, in the same bubble. To this extent, one usually takes Wigner to live in a different bubble than the friend because the latter is enclosed in a fully isolated laboratory. Yet, a bubble is notably an information-theoretic concept and, as such, is fundamentally independent of space-time location \cite{pienaar2023single}, relying exclusively on the relevant information that is, in principle, retrievable. 
Thus, in general, the existence of the bubble is not compromised even if information is exchanged between the parties, provided that this information is uncorrelated with the outcomes of the quantum measurements. (An asymmetry in the flow of outcome-related information from the friend to Wigner, as opposed to from Wigner to the friend, will be discussed later.)

The already mentioned no-go theorems for extended Wigner's friend scenarios \cite{Brukner2015, Frauchiger2018, Brukner2018,  bong2020strong} formally state that, under certain assumptions, there is no joint probability for the results obtained by the friend and Wigner such that the respective probabilities for their results can be understood as marginals of it.
The bubble view is a way to resolve the ambiguity of the state assignment, insofar as a state can be considered \textit{objective only relatively to a specified bubble}, i.e., each and every hypothetical observer within the same bubble will be entitled to assign the same state within their respective bubble. However, the states will in general be different for different bubbles.

The strategies for rational predictions are less clear when the observers can “switch" bubbles, for example, when a measurement is performed in another bubble, but the player finds a way to verify the outcomes of this measurement and thus their previous predictions.
In such cases, if an observer who resides in one bubble has access to a quantum state assignment from another bubble, should they adopt it for the purpose of predicting the probabilities for the occurrence of the measurement outcomes? Imagine, for example, that the friend knows the state in which Wigner describes her laboratory, and she also knows which measurement of her laboratory Wigner will make next. Should she condition her predictions solely on the knowledge of the measurement results she observed in her lab, or on Wigner's knowledge.
And given that the future measurement will affect her memory, will she even be able to evaluate her previous prediction after the measurement has been completed? 


The first step towards investigating situation when an observer may “switch" the bubbles was taken in Ref.~\cite{baumann2020wigner}, but only for the case where the friend was placed in the situation. Here we aim at the situations more symmetrical for the two observers: we propose a game (in Sect. \ref{sec:3}) when the knowledge of the two observers is not only maximally incompatible (i.e. when the two measurements in the respective laboratories are maximally non-commutative), but partially overlapping, thus covering the whole spectrum from full compatibility to maximal incompatibility. In that sense our work amounts to the task of finding information-theoretic conditions that allow to univocally decide which state assignment is rational to adopt in cases when an observer can ``switch" from one bubble to another and when some information is allowed to be exchanged between bubbles.

\section{Communicating “bubbles"}

Let us consider the qubit system (S) and the friend's (F) and Wigner's (W) environments or laboratories. At the beginning the situations of the friend and of Wigner are fully symmetrical and similar to the one experienced by our experimental colleagues in everyday practice. In particular, we assume that they each possess reliable experimental devices and act as rational agents. Moreover, the interaction between the quantum system and the environment in each laboratory results in environmentally-induced decoherence and a \textit{preferred} environmental basis in which measurement outcomes are redundantly recorded~\cite{Zurek2003,zurek2009quantum, Zurek2018, schlosshauer2019quantum}. The records may include “clicks" in a detector, laboratory notes or even the observer's memory. Nevertheless, for the purposes of our argument, we will assume that the preferred bases of the friend's and Wigner's environments are not identical, and that there is a “boundary" between two bubbles, in which a unitary transformation maps one basis into the other.

We assume that (macroscopic) degrees of freedom of the environments, among other states, can take two distinct states, $|\phi_0\rangle_F$ or $|\phi_1\rangle_F$ in the preferred basis of friend's and $|\omega_0\rangle_W$ or $|\omega_1\rangle_W$ in the preferred basis of Wigner's laboratory. The preferred bases in the respective laboratories will contain these pairs of states. The probabilities for finding these pairs of states will be denoted as $p^F_{\phi_i}$ $(i=0,1)$ in the friend's and $p^W_{\omega_j}$ $(j=0,1)$ in Wigner's laboratories.

We now consider Wigner's friend though experiment. Consider the following initial state 
\begin{equation}
\label{psi0}
    \ket{\Psi_0}=\ket{\psi}_S\ket{\phi_0}_F\ket{\omega_0}_W,
\end{equation}
where $\ket{\psi}_S$ is a state  of the (qubit) system and $\ket{\phi_0}_F$, $\ket{\omega_0}_W$ are no further specified “ready" states of the friend's and Wigner's environment, respectively. At this stage, the state assignment could be one upon which both the friend and Wigner agree. We assume for simplicity
$\ket{\psi}_S=\frac{1}{\sqrt{2}}(\ket{\uparrow}_S+\ket{\downarrow}_S)$  {where $\{\ket{\uparrow}_S,\ket{\downarrow}_S \}$ is the orthonormal basis of the qubit in  which the friend measures, say the $z$-direction of the spin}. 

The friend then performs her measurement. In Wigner's description, this corresponds to the application of a unitary operator $U_\mathrm{meas}$ that correlates the degrees of freedom of the system and of friend's environment (which includes her memory). We assume, for simplicity, that $U_\mathrm{meas}$ implements a quantum controlled-NOT operation such that if the spin is in state $\ket{\uparrow}_S$, the environmental state $\ket{\phi_0}_F$ remains unchanged, and if it is $\ket{\downarrow}_S$, it flips the state to $\ket{\phi_1}_F$. After the measurement the overall state of the system and the friend's environment becomes perfectly correlated:
\begin{equation}
\label{psi1}
     \ket{\Psi_1^W}= U_\mathrm{meas} \ket{\Psi_0}= \frac{1}{\sqrt{2}}\Big(\ket{\uparrow}_S\ket{\phi_0}_F+\ket{\downarrow}_S\ket{\phi_1}_F \Big)\ket{\omega_0}_W,
\end{equation}
where the superscript $W$ indicates that this is the state assigned by Wigner. Note that the presence of Wigner's environment doesn't play any role yet at this stage since he has not performed any measurement.

It is important to note that we are assuming that there are degrees of freedom, not explicitly denoted in Eq.~(\ref{psi1}), in the friend's lab that do not get correlated with the outcome. It is these degrees of freedom that can and will be freely used in communication with Wigner's bubble without decohering the state. 
 
On the other hand, the friend performing the measurement on the qubit system enforces the measurement (state-update) postulate by the application of projectors $\Pi_\uparrow = \ket{\uparrow}_S\bra{\uparrow}$ and $\Pi_\downarrow = \ket{\downarrow}_S\bra{\downarrow}$ over $\ket{\psi_s}$ in Eq.~\eqref{psi0}. An equivalent perspective on the measurement process for the friend is to first adopt the state assignment~(\ref{psi1}) for the qubit and her environment, while concurrently applying the projector $\Pi^F_0 = \ket{\phi_0}_F\bra{\phi_0}$ and $\Pi^F_1 = \ket{\phi_1}_F\bra{\phi_1}$ over $\ket{\Psi^W_1}$ in Eq.~\eqref{psi1}.\footnote{Provided that the friend can only read out the states in the preferred basis of the environment, two friend's descriptions are possible. Either by applying the state update rule to the quantum system in state~\eqref{psi0} or by applying the state update rule to his environment in state~~\eqref{psi1} are equivalent. Indeed, when performing experiments, our experimental colleagues apply the state update rule to the measured system. Nevertheless, they may still employ description~~\eqref{psi1} to analyse the decoherence process induced by the measurement. Note that in either case, the state of the Wigner's environment is irrelevant and can be traced out.} This leads to  classically correlating the friend environment to the measurement result in the following states, depending on the friend's measurement outcome:

\begin{equation}
\label{measm}
   \ket{\Psi_1^F}= 
\begin{cases}
   \ket{\uparrow}_S\ket{\phi_0}_F\ket{\omega_0}_W  ,& \text{for outcome } \phi_0\\\vspace{0.1cm}\\
    \ket{\downarrow}_S\ket{\phi_1}_F\ket{\omega_0}_W,              & \text{for outcome } \phi_1 \end{cases}
\end{equation}
each of them occurring with probabilities $p^F_{\phi_0} = p^F_{\phi_1} = 1/2$. Here the superscript $F$ stands for the friend's state assignment and Wigner's environment is added just for completeness but is  unaccessible to her. 

Let us consider now that a hypothetical physical process at the boundary of the two labs couples the two environments as illustrated in Fig. 1. This coupling is described by a 
unitary operator $U_\mathrm{int}$ acting over the system and the two environments and  both observers know this unitary. The unitary is defined through its action on the only states on which it acts non-trivially:

\begin{equation}
\label{collmeas1}
\begin{split}
     {U_{\text{int}}} \ket{\uparrow}_S
     \ket{\phi_0}_F\ket{\omega_0}_W     &~ {=}~  \frac{1}{\sqrt2} (\ket{\Phi^+}_{SF}    \ket{\omega_0}_W+ |\Phi^-\rangle_{SF} \ket{\omega_1}_W), \\ 
      {U_{\text{int}}} \ket{\downarrow}_S
     \ket{\phi_1}_F\ket{\omega_0}_W     &~ {=}~  \frac{1}{\sqrt2} (\ket{\Phi^+}_{SF}    \ket{\omega_0}_W - |\Phi^-\rangle_{SF} \ket{\omega_1}_W),
     \end{split}
\end{equation}
where $\ket{\Phi^\pm}_{SF} = (\ket{\uparrow}_S\ket{\phi_0}_F \pm \ket{\downarrow}_S\ket{\phi_1}_F)/\sqrt2$ are two of the standard Bell's states over the system's and friend's Hilbert spaces. 
Therefore, by applying $U_{\text{int}}$ to Eq.~\eqref{psi1} we obtain for Wigner's description:
\begin{align}
 \label{psiU2} \ket{\Psi_2^W} =
   U_{\text{int}} \ket{\Psi_1^W} = \ket{\Phi^+}_{SF}\ket{\omega_0}_W.
\end{align}

\begin{figure}[ht]
\includegraphics[width=\linewidth]{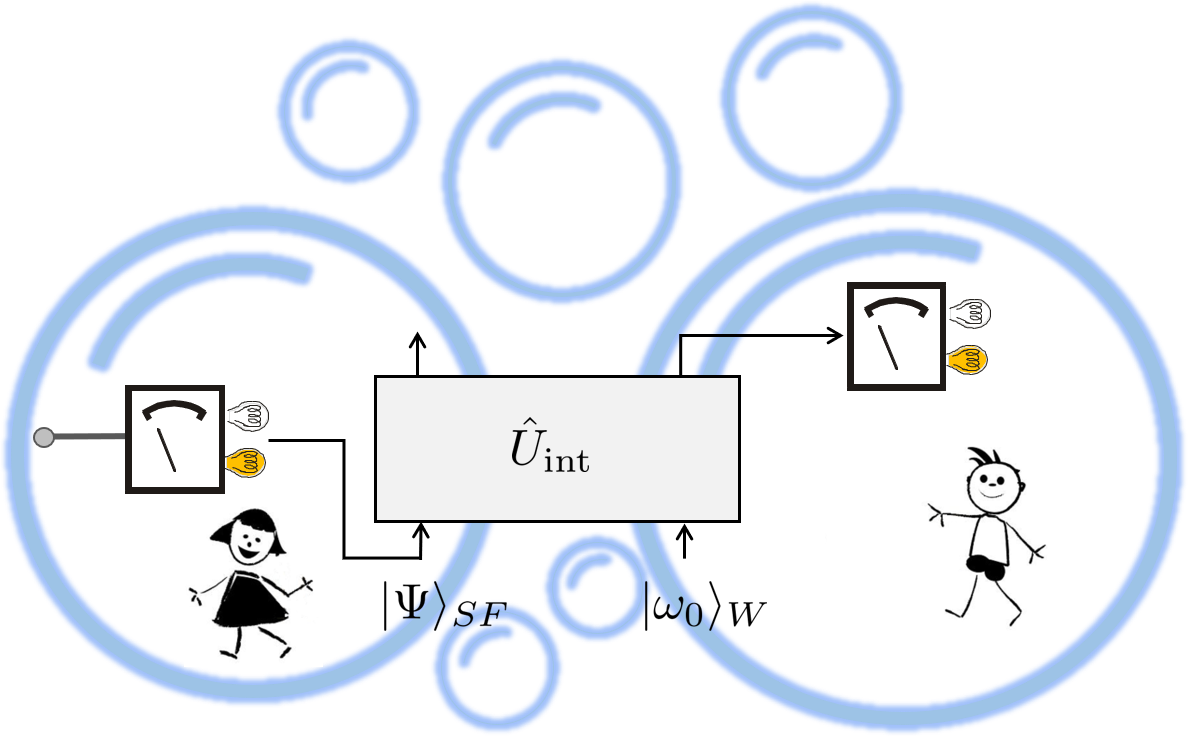}
\caption{Wigner and his friend assign in general different states to the same physical systems, i.e., they live in different “bubbles". In the friend's bubble a measurement of a qubit is performed and an outcome is obtained. We assume that there exists an interface at the “boundary" of the two bubbles where the environmental states of the friend's and Wigner's bubble unitarily interact. Depending on the unitary $\hat{U}_{\text{int}}$, information about the results of the friend's measurements may or may not be encoded in the Wigner's environment, making the descriptions of the friend and the Wigner compatible (hence merging the two bubbles in one) or not. In the extreme case of the unitary~(\ref{collmeas1}),
there is no information about the friend's outcome in Wigner's environment.}
\end{figure}

 The  unitary operation does not affect the initial correlations between the state of the qubit and the friend's environment. Hence, looking at his environmental degree of freedom, Wigner can only access the information about the system and the friend being perfectly correlated in state $\ket{\Phi^+}$, while the actual information about “which-outcome" $\uparrow$ or $\downarrow$ of the qubit measurement is inaccessible. In Appendix A we consider more general unitaries where partial information about friend's measurement outcome is leaked into Wigner's environment.

On the other hand, applying the unitary $U_{\text{int}}$ to the friend's post-measurement states in Eq.~\eqref{measm} we get the states:
\begin{equation}
\label{measU2} \ket{\Psi_2^F}=
\begin{cases}
   \frac{1}{\sqrt2} (\ket{\Phi^+}_{SF}    \ket{\omega_0}_W+ |\Phi^-\rangle_{SF} \ket{\omega_1}_W), \\\\ 
   \frac{1}{\sqrt2} (\ket{\Phi^+}_{SF}    \ket{\omega_0}_W - |\Phi^-\rangle_{SF} \ket{\omega_1}_W)
\end{cases}
\end{equation}
associated respectively to $\phi_0$ and $\phi_1$ results in Eq. \eqref{measm}. 
The above state can be equivalently written as
\begin{equation}
\label{measU22} \ket{\Psi_2^F}=
\begin{cases}
  \frac{1}{\sqrt{2}}( \ket{\uparrow}_S\ket{\phi_0}_F \ket{\omega_+}_W + \ket{\downarrow}_S\ket{\phi_1}_F \ket{\omega_-}_W ), \\\\ 
  \frac{1}{\sqrt{2}}( \ket{\uparrow}_S\ket{\phi_0}_F \ket{\omega_-}_W + \ket{\downarrow}_S\ket{\phi_1}_F \ket{\omega_+}_W ),
\end{cases}
\end{equation}
with $\ket{\omega_{\pm}}_W = (\ket{\omega_0}_W \pm \ket{\omega_1}_W)/\sqrt{2}$. From Eq.~(\ref{measU2}), it can be seen that Wigner's environmental states $\{\ket{\omega_0}_W, \ket{\omega_1}_W$ \} are perfectly correlated with the Bell states of the system and the friend's environment. This implies that, even in the friend's state,  no information about the outcome of the friend's measurement can be inferred from the preferred environmental basis in Wigner's laboratory. Moreover, from Eq.~(\ref{measU22}), it can be seen that for any of the two outcomes of the qubit measurement and following the action of $U_\mathrm{int}$, the qubit system and the friend's laboratory become entangled in a state with two ``branches''. This demonstrates that the correlation between the spin and the laboratory's record is still maintained after the unitary interaction.

\color{black}

For the purpose of the game, we will assume that after the first measurement of the qubit by the friend and the action of the unitary at the interface, one of two possible measurements will be performed next. Either the friend will repeat the same measurement ($\text{M}_F$) of her qubit, measuring again the spin along the $z$ direction, or alternatively, Wigner will perform a measurement ($M_W$) of the entire friend's laboratory in the preferred basis of his environment, i.e., using projectors $\Pi_0^W = \ket{\omega_0}\bra{\omega_0}_W$ and $\Pi_1^W = \ket{\omega_1}\bra{\omega_1}_W$. In the game, the two observers will be required to provide their predictions regarding the outcomes of one or the other measurement.  
\section{The “Bubble Switching Game"}
\subsection{The rules of the game} \label{sec:3}

Next we will introduce the rules of the game. The game involves two players, Wigner (W) and the friend (F), as well as two referees,\footnote{To avoid unnecessary complications in the notation, we treat these referees as co-observers who share access to the relevant outcomes within their bubble. While they are not modeled explicitly, we assume they register and agree with the measurement results of the agents in their respective laboratories.} R$_W$ and R$_F$, located in Wigner's and the friend's bubbles, respectively. 

\begin{mdframed}

\begin{enumerate}

    \item Each player knows the state in their respective bubble. In particular, F knows the definite outcome of the first measurement (and can also compute state $|\Psi_2^F\rangle$), and  W knows $|\Psi_2^W\rangle$.
    \item The next measurement, either M$^F$ or M$^W$, is chosen at random.
    \item The players submit their predictions regarding the probabilities of measurement outcomes to the referee in the bubble where the measurement will be conducted, to R$^F$ in case of M$^F$, or R$^W$ in case of M$^W$.
    \item 
    The measurement is performed. 
    \item The referee, situated within the bubble in which the measurement was performed, awards the player whose prediction is confirmed by the measurement.

\end{enumerate}

\end{mdframed}
It is important to note that all forms of communication, including the submission of predictions and the sharing of random choices regarding the next measurement, will not contain any information about the outcome of the first friend's measurement and will not affect the states as assigned in the bubbles. In this sense, a bubble is an information theoretic concept, rather linked to the kind of information in principle available than to an amount of leaked non-relevant information (i.e., there is no need of complete isolation between two regions for them to be two bubbles; see Sect. \ref{bubbles}).

We will now proceed to comment on the rules of the game. Note that the players do not necessarily need to rely on the honesty of the referees in step 5 when evaluating their predictions. Instead, after each run of the game, they can inspect their previous predictions in form of permanent records (e.g., signed pieces of paper on which their prediction is written) that they handed to the responsible referee in step 3. Also, for case $M_W$ the friend's memory will be subjected to a measurement thereby erasing that (tiny) part of their memory which recorded the outcome of the previous measurement. This may give rise to the concern that the “forgetful" observer may not agree that they ever made any prediction, or at least not the one which they will be faced with at the end of the game. However, since the players may still be presented with a record of their previous predictions --the signed piece of paper with their signature-- this may provide them with sufficient evidence that the prediction was indeed made by them.  

Finally, in order to avoid potential difficulties arising from the so-called \textit{self-reference problem} \cite{breuer1995impossibility}, namely, the circumstance that the friend might have to measure her own state, let us append to the already considered scenario a further auxiliary memory system (A) of the same dimension than $F$, prepared in an initial “ready" state $\ket{R}_A$. 
Applying then a unitary swap operation, $U_\mathrm{SWAP}$ acting over friend and the auxiliary system, the memory of the friend is coherently transferred to the auxiliary system (such that it swaps the indices of the friend and the auxiliary system). Under the action of $U_\mathrm{SWAP}$, the states in Eq.~\eqref{psiU2} and~\eqref{measU2} transform as
\begin{align}
    U_\mathrm{SWAP} \ket{R}_A \ket{\Psi^{W}_2} &=  \ket{\Psi^{W}_3} \ket{R}_F, \\
   U_\mathrm{SWAP} \ket{R}_A \ket{\Psi^{F}_2}  &= \ket{\Psi^{F}_3} \ket{R}_F,
\end{align} 
respectively. This SWAP operation can be performed by Wigner himself or, if the friend does not trust him, by a hypothetical machine that the friend programs beforehand for carrying out this task. In such a way, the friend would be able to measure a state containing a record of what she observed without having to measure her own state. By applying the SWAP operation, the content of the friend's memory will be erased, and the comments previously made about the loss of her memory can be applied to this step as well. Note that Wigner's memory is not affected by this operation.

In what follows we will demonstrate that, in accordance with the established rules of the game, players will be led to conclude that it is preferential to adopt the state assignment of the other bubble when the measurement is performed in that other bubble, while maintaining their own for all measurements performed within their own bubble. In certain instances (for the friend), the state assignment from the other bubble may be accessible in principle, whereas in others, it may not be (Wigner). 

\subsection{Measurement in Wigner's bubble} \label{sec:measW}

Let us suppose that the random choice determines that the measurement \( M_W \), in which Wigner measures in the preferred basis of his environment \(\{\ket{\omega_0}_W, \ket{\omega_1}_W\}\), should be made next.
 On the basis of his state assignment~(\ref{psiU2}), Wigner makes a prediction that the two outcomes will occur with probabilities $p^W_{\omega_0}=1, p^W_{\omega_1}=0$. He then hands this prediction to referee R$_W$. The aforementioned prediction is equally applicable also in the case when the SWAP operation is applied prior to the $M_W$ measurement. 

On the other hand, when applying the SWAP operation the state of the friend is just Eq.~\eqref{measU2} where index F have been replaced by A. 
She predicts that probabilities for observing one or the other outcome of $M_W$ are $p^F_{\omega_0}=p^F_{\omega_1}=1/2$  (not to be confused with the friend's probabilities of the first qubit measurement). These predicted values for the probabilities  can be encoded in a degree of freedom of her laboratory that did not get correlated to the outcome of the previous qubit measurement. Since the message containing the prediction also carries no information about the outcome of friend's qubit measurement, it factorises out from her laboratory and can be transmitted to Wigner's laboratory to the responsible referee R$_W$ as illustrated in Fig.~2.

\begin{figure}[ht]
\includegraphics[width=\linewidth]{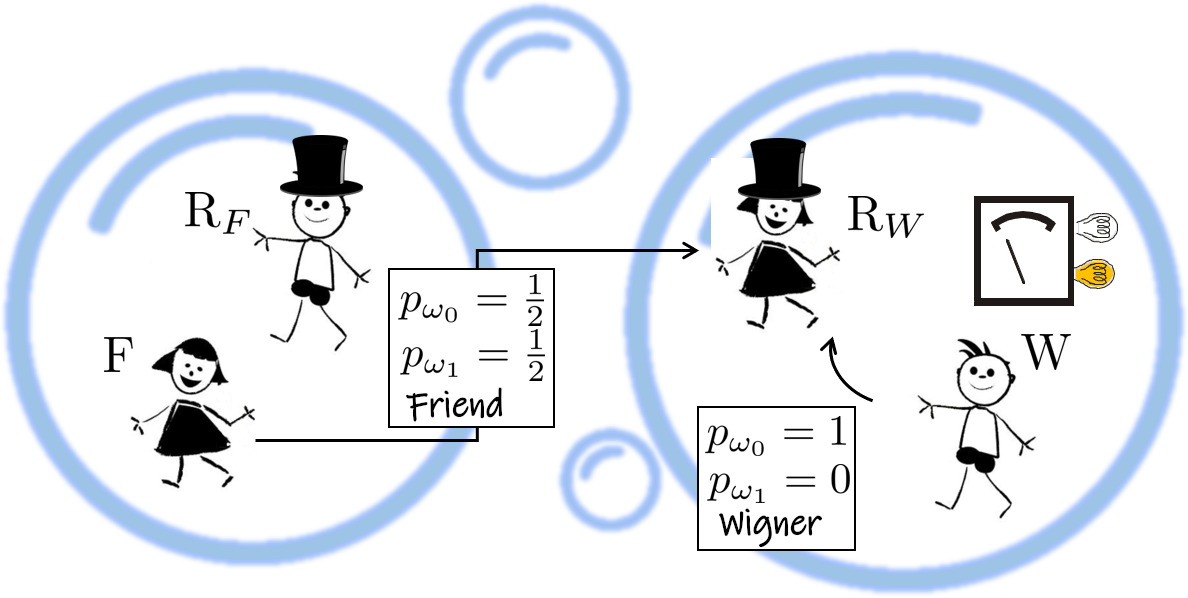}
\caption{The schematic illustration of the game protocol in the case that measurement $M_W$ in Wigner's bubble is chosen to be performed. On the basis of quantum state assignments in their respective laboratories, the two players make their predictions and submit them to the responsible referee R$_W$ in Wigner's bubble. The probabilistic prediction of the friend contains no information about the outcome of the qubit observed in her laboratory. Consequently, its communication to the referee has no effect on the state of the laboratory as described by Wigner. The outcome of the $M_W$ measurement is in accordance with the prediction of Wigner, and in repeated trials, a discrepancy from the prediction of the friend can be observed.}
\end{figure}

The measurement is then performed in Wigner's bubble and the result $\omega_0$ is observed. Since the friend and Wigner predictions differ in the estimated probabilities, one would have to perform a sufficient number of runs to verify the difference between the two predictions with a certain degree of confidence, always keeping the records of both the measurement outcomes and the two predictions of the previous runs. The observed relative frequency will eventually show a significant deviation from the friend's prediction as given, e.g., by Wald's sequential probability ratio test~\cite{Wald1945}.\footnote{One can also construct a scenario in which Wigner's assigns probability one to a specific outcome, while the friend assigns a probability that is arbitrarily close to zero, see Ref.~\cite{baumann2020wigner}. This would render the argument arbitrarily close to a deterministic one.} and consequently the referee R$_W$ will award Wigner. Interestingly, this situation is maintained even in the cases where partial information about the qubit measurement outcomes are leaked into Wigner's environment, as explicitly discussed in Appendix~\ref{appB}.

Note that the measurement procedure affects the friend's memory in that the content of the memory that contained the information about the outcome have been swapped with the ancilla state. While this allowed the friend to avoid the problem of self-reference, her memory of the outcome was lost in the process. One might ask, then, whether the  loss of this part of the memory would alter the entire game to such an extent that the friend's identity as an observer would be lost, and she could no longer be regarded as the same player who started the game. We leave this 
point for the discussion, and remind the reader at this point that there is still a record (possibly with her signature) of the friend's previous prediction, which she can inspect and accept as valid evidence of her previous prediction. 

The result of the game shows us that the friend could have made a more accurate prediction if she adopted the Wigner state assignment whenever the measurement $M_W$ is chosen. In fact, she \textit{can} do this, since there are no fundamental restrictions as to why this information should be inaccessible to her. In fact, Wigner could simply send this information to her bubble, or the friend herself could first construct the experimental situation from Wigner's perspective and then enter the sealed laboratory that becomes her bubble.

\subsection{Measurement in the friend's bubble} \label{sec:measF}


Next we consider the case where the decision is made to take the measurement $M_F$. This means that the measurement of the qubit along the $z$-direction in the friend bubble using projectors $\Pi_\uparrow$ and $\Pi_\downarrow$ is repeated. The responsible referee is R$_F$. The game is illustrated in Fig.~3.

\begin{figure}[ht]
\includegraphics[width=\linewidth]{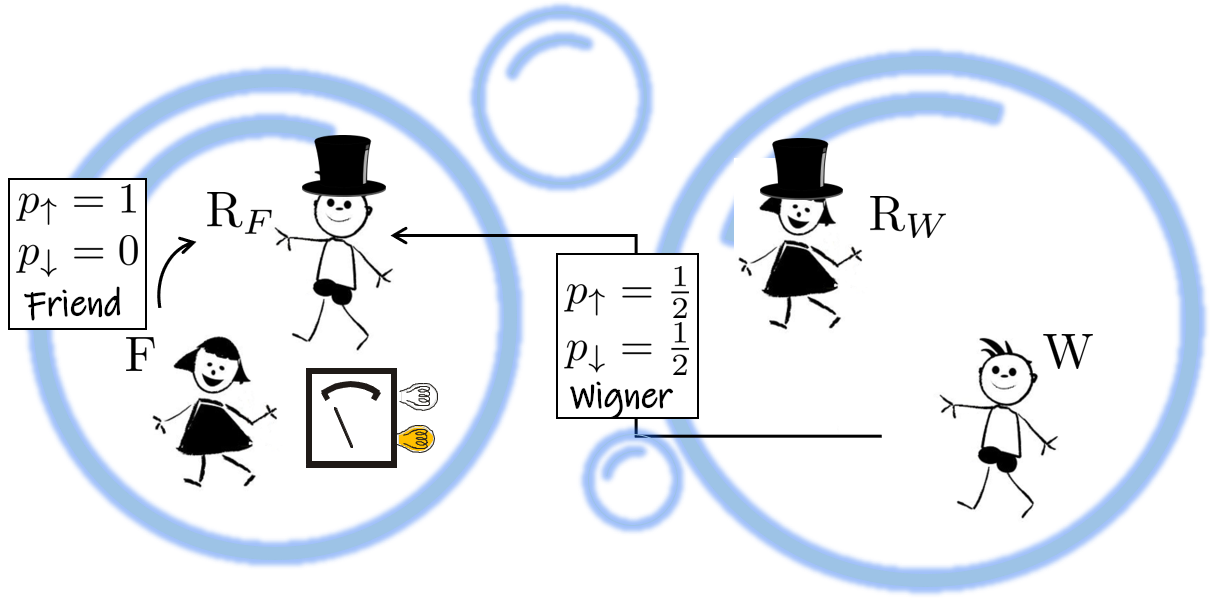}
\caption{The schematic illustration of the game protocol in the case that measurement $M_F$ in the friend’s bubble is chosen to be performed. The two players make their predictions on the basis of their state assignments and submit them to the responsible referee R$_F$ in the friend’s bubble. The prediction of Wigner can be sent inside the friend's laboratory with no additional effects. The outcome of the measurement is consistent with the prediction of the friend, and in repeated trials, a discrepancy between the observed relative frequencies and the prediction of Wigner can be observed.}
\end{figure}


Which state is the friend going to use as a basis for her prediction? After the action of the unitary interaction $U_\text{int}$ the friend's state becomes Eq.~(\ref{measU2}). It is important to note that this is \textit{not} the state that the friend would normally adopt in order to make predictions about measurements in her laboratory. Rather, it is the state that the friend estimates she should use for predicting results of measurement in Wigner's bubble, such as $M_W$, after the action of the unitary and conditional on knowledge of her measurement outcome. In fact, as discussed below Eq.~(\ref{measU22}), from the perspective of the friend, the state of her memory (or of the ancilla after the SWAP) remains in a definite state perfectly correlated with that of the spin even after the action of the interaction unitary. For all subsequent measurements in her lab, all that matters for predicting future results is the last record in her lab. And that will be either definitely $\uparrow$ or definitely $\downarrow$.\footnote{It may happen, however, that the action of the unitary induces a flip in the friend's memory, but even then the friend will still have \textit{a} definite record of the first measurement in her bubble. All her subsequent predictions can then be conditioned on \textit{that} record. Alternatively, in ``many-worlds" terms, one might say that unitary evolution ``splits'' the world into two branches (as given in each of the two lines of Eq. \eqref{measU22}), so that in each branch a friend sees a definite (and different) outcome. Also in this view, in order to predict the outcome of any future measurement within her branch, the friend should condition her prediction only on the last record in her branch.} Thus, the friend's description of the relevant degrees of freedom for predicting future measurement results in her laboratory remains to be given by state~(\ref{measm}), even after the action of the unitary interaction.  


If the friend repeats the same measurement, she can use an additional degree of freedom F' in her laboratory, which was initially sufficiently isolated that it did not get correlated with the previous result, and can be used to store the outcome of the repeated measurement. To formally describe it, one could start from state (\ref{measU22}) obtained after the unitary interaction, add the degree of freedom F' and apply the quantum controlled-NOT operation acting between the spin and degree F', as a purified description of the repeated measurement.  The final state would be similar to Eq.~(\ref{measU22}), with the difference that states $\ket{\phi_0}_F$ and $\ket{\phi_1}_F$ of a single record in the two branches would be replaced by states $\ket{\phi_0}_F \ket{\phi_0}_{F'}$ and $\ket{\phi_1}_F \ket{\phi_1}_{F'}$ of two copies of the record. However, as we have argued, only that part of the state that features the correlations between the spin and laboratory degrees of freedom are relevant for predicting the outcomes of future measurements in the friend's laboratory. The friend could just as well discard the action of the unitary interaction and describe her repetition of the measurement simply by starting from Eq.~(\ref{measm}), adding the degree of freedom F', and applying the quantum controlled NOT to the spin and F'.

Hence, if she repeats the same measurement, she will always confirm the record of the previous measurement result (which she had after the action of the unitary): she finds the outcome $\uparrow$ with probability $1$ and $\downarrow$ with probability $0$ if the record of her previous observation was $\uparrow$, and similarly she finds $\downarrow$ with probability $1$ and $\uparrow$ with probability $0$ if the record of her first measurement was $\downarrow$. 
\color{black}
Hence, she predicts $p^F_{\phi_0}=1$ (or 0), $p^F_{\phi_1}=0$ (or 1) and gives this prediction to referee R$_{F}$. Formally, the state assigned by the friend after the repeated measurement is given by: 
\begin{equation}
\label{meas}
   \ket{\Psi^F_4}= 
\begin{cases}
   \ket{\uparrow}_S\ket{\phi_0}_F\ket{\phi_0}_{F'}\ket{\omega_0}_W  & \text{for outcome, } \phi_0\\\vspace{0.1cm}\\
    \ket{\downarrow}_S\ket{\phi_1}_F\ket{\phi_1}_{F'}\ket{\omega_0}_W              & \text{for outcome } \phi_1. \end{cases}
\end{equation}
From Wigner's point of view, a repetition of the friend's measurement is a repeated application of the measurement unitary $U_{\text{meas}}$, which acts as a quantum controlled-NOT operation between the system and the additional degree of freedom $F'$. This implies that after the repeated measurement the state assigned by Wigner is 
\begin{align}
    \ket{\Psi^W_4} &=
    (\hat{U}_{\mathrm{meas}} \otimes \mathds{1}_{F}) \ket{\Phi^+}_{SF}\ket{\phi_0}_{F^\prime}|\omega_0\rangle_W\\
    & = \ket{\mathrm{GHZ}}_{SFF'} \ket{\omega_0}_W \nonumber,
\end{align}
where
\begin{equation}      \ket{\mathrm{GHZ}}= \frac{1}{\sqrt{2}}\Big(\ket{\uparrow}_S\ket{\phi_0}_F\ket{\phi_0}_{F'}+\ket{\downarrow}_S\ket{\phi_1}_F\ket{\phi_1}_{F'} \Big)
\end{equation}
is the Greenberger-Horne-Zeilinger (GHZ) state. The probabilities for the outcomes $\uparrow$ and $\downarrow$ are both $p^W_{\phi_0}=p^W_{\phi_1}=$1/2 in this state. Wigner sends this prediction to referee R$_F$ inside the friend's bubble. As in the case of the measurement in Wigner's bubble, the friend's and Wigner's predictions differ in their estimated probabilities, which can be compared to a desired degree of confidence if a sufficient number of runs were made, always keeping track of both the measurement results and the two predictions from the previous runs. The observed relative frequency will eventually show a significant deviation from Wigner's prediction and the friend will be awarded by the referee. Similar conclusions are obtained in the case of partial leakage of which outcome information into Wigner's environment (see Appendix~\ref{appB}).

We conclude that the friend's state assignment was more accurate than Wigner's for a measurement carried out in her laboratory. However, unlike the case of the $M_W$ measurement, where the friend could adopt Wigner's prediction before the measurement was carried out and thus estimate the results of the measurement as well as Wigner, this is not true in the present case since any leakage of the information about the result of the first friend's measurement to Wigner would correspondingly collapse his state.

\section{(In)compatibility between different bubbles}

In the case where the information about which outcome the friend has observed cannot be unambiguously retrieved from the Wigner's environment, Wigner and the friend give different, irreconcilable state assignments. None of them is ``right" and the other ``wrong". They live in different “bubbles"~\cite{cavalcanti2021view} and each of the state assignment is “right" relative to the measurements performed in their respective bubbles.
In these cases both Wigner and her friend might be tempted to think that the other party is making a wrong assignment. On the one hand, Wigner may think that the operation performed by the friend was not actually a ``proper measurement", since, following his description, unambiguous information about the outcomes is not detectable anywhere in his bubble. On the other hand, the friend---having evidence that a genuine measurement has taken place---may regard Wigner’s unitary description as inadequate, since it fails to account for the for-all-practical-purposes irreversible imprint of the measurement outcome within her bubble. In any case, the assumption that no observer is preferred and that observers in other bubbles may experience definite outcomes, just as they do in their own, may prompt them to consider the possibility of bubbles with fundamentally incompatible outcomes.

On the contrary, as explicitly shown in Appendix~\ref{appB}, in the special case in which information about “which-outcome" is completely leaked into Wigner's environment (i.e. each outcome is correlated to an accessible distinguishable state of the environment), the measurement update rule can be safely applied by both observers. One can say that if both observers share the information about the outcomes then they live in the same “bubble". However, it is enough to have an arbitrarily small uncertainty in Wigner's environmental states about the result of the qubit measurement in friend's lab 
to create an (in principle) detectable incompatibility between Wigner and the friend's state assignment. 
Interestingly, the number of runs needed to determine the validity of Wigner (friend) predictions versus the alternative one is a monotonically increasing function of the encoding parameter representing the amount of leaked information (see Appendix~\ref{appB}): it increases as the amount of information leaked into Wigner's environment increases and diverges when the two descriptions are compatible. 

It is important to note that the common assumption that the friend must be in a perfectly closed lab is unwarranted. The key distinction lies in the type of information that leaks outside: any non-substantial information about the measurement procedure---such as the fact that the system and the friend are entangled from Wigner’s perspective---or unrelated topics like the weather forecast, a film critique, etc., can be communicated between Wigner and the friend without affecting their state assignments. Similarly, Wigner and his friend can exchange their predictions, share the random choice determining the subsequent measurement, or use communication to verify that their descriptions are incompatible.

Finally, one can consider the possibility of a hierarchy of numerous bubbles, each of which can be measured from the “higher" ones in the hierarchy. In that case, we may conclude that a complete description of each observer's physical situation would require a corresponding hierarchy of joint state assignments: the friend's state for measurements in the friend's bubble, Wigner's state for measurements in Wigner's bubble, and so on. It is important to reiterate that an observer cannot have the state assignment of any bubbles “below" themselves in the hierarchy without decohering them. For instance, Wigner cannot have the state assignment of the friend without collapsing it. But it is this asymmetry that allows the “lower" observer in the hierarchy to be more successful in certain tasks than the “higher" level observer.



\section{Conclusions}

According to the fundamental principles of textbook quantum mechanics, an observer may apply two distinct types of evolution to a quantum state, contingent upon whether the evolution takes place between measurements or following a measurement.
 In the former case, a unitary transformation is applied, whereas in the latter, projections are employed (in the ideal case). However, textbooks are silent on the conditions that specify which evolution should be applied in a concrete situation. This ambiguity can be attributed to a fundamental issue in defining the characteristics distinguishing between laboratory devices that serve as measuring instruments and those that implement unitaries -- an issue known as “the quantum measurement problem".

In this work, we did not address the measurement problem. Rather, we acknowledge that through everyday praxis, an observer gains experience with various devices and can distinguish between those that function as measurement apparatuses and those that implement unitary evolutions. This is the starting point for our discussion. Nevertheless, Wigner’s thought experiment and its modern extensions present an additional challenge to the consistency of the textbook quantum mechanics. This is because different observers may assign different quantum states to the same system. One possible way to overcome this challenge is to adopt the view that the fundamental concepts of quantum mechanics, such as quantum state, unitary evolution, or updated state, can only be defined relative to a bubble. Within a single bubble, all hypothetical observers may share their knowledge, thereby establishing intersubjectivity within the bubble. Since a bubble is defined only in terms of the (kind of) information that is in principle available in a certain region, we call this “relative objectivity". In fact, according to this view, there is generally no universally correct state. However, once the physical conditions are specified (i.e., the bubble is defined), the state is determined. This is reminiscent of relativity theory, where for example simultaneity between space-like separated events is not absolutely defined, but this relation is fixed within any chosen foliation \cite{sudbery2017single}. It would be desirable to find analogous ways to formally transform the description of a bubble into another by developing the analogue of  Lorentz transformations.

However, we believe that a transformation from one observer’s perspective to another---such as from the friend to Wigner---may not, in general, be representable by a unitary operation (as in the framework of quantum reference frame transformations \cite{Giacomini2017}). In fact, from the friend's perspective, the measurement has taken place, resulting in two or more classically distinguishable outcomes—i.e., distinct states corresponding to the possible measurement results. From Wigner’s perspective, however, the friend and the measured system remain in a single, entangled state. Crucially, since unitary transformations are bijective, they cannot map multiple classical states corresponding to the friend’s possible distinct experiences onto a single entangled state as described by Wigner. This suggests that no single unitary can bridge the two perspectives.

We examined scenarios in which observers can transition between bubbles, gaining (partial) knowledge of the state of another bubble. The observers are then tasked with making predictions about measurements that will be made in or on their own bubbles, including their own memory. Our findings highlight the need for a refinement of the concept of objectivity relative to the bubbles. We demonstrated that, under certain circumstances, observers are entitled to adopt the state assignment from other bubbles where measurements are performed and make predictions accordingly. A true novelty is that, in different cases, both Wigner and the friend can achieve this. This symmetrizes the scenario and challenges the view that Wigner, as a superobserver, always possesses more predictive power than the friend.



Finally, we comment on the memory loss of the aforementioned friend whose laboratory, including his memory, underwent the measurement procedure. It is sometimes argued that, because of the loss of memory, the person should not be considered the same throughout the experiment (see \cite{guerin2021no} and references therein for a discussion). In such a case, the individual would not be considered the same “friend" or “player" in our game. Consequently, they might not accept that they made a prediction in the past that led to them losing the game. 
The conventional notion of personal identity is defined by the distinctive ways in which an individual defines themselves, encompassing aspects such as self-image, cultural references, political credos, or other interests. Since the loss of memory in the measurement solely affects the recollection of the outcome of the very last measurement, these fundamental aspects of identity are expected to remain unaltered. Moreover, it is not uncommon for individuals to forget events from the past. However, when presented with compelling evidence that an event occurred in the past, they tend to accept that it did occur and that they simply forgot about it. Similarly, in our game, after players are presented with the signed paper containing their prediction, they may accept it as legitimate. In the end, one might approach the problem of identity in the present case pragmatically: Would \textit{you} enter the game under the risk of losing the memory of one bit of information, if the award were one million coins? 


\section*{Acknowledgements} \label{sec:acknowledgements}
 FDS and {\v C}B acknowledge the discussion with Veronika Baumann. This research was funded in whole or in part by the Austrian Science Fund (FWF) 10.55776/COE1 (Quantum Science Austria), [10.55776/F71] (BeyondC) and [10.55776/RG3] (Reseacrh Group 3). For Open Access purposes, the authors have applied a CC BY public copyright license to any author accepted manuscript version arising from this submission. This publication was made possible through the financial support of the ID 62312 grant from the John Templeton Foundation, as part of The Quantum Information Structure of Spacetime (QISS) Project (qiss.fr). The opinions expressed in this publication are those of the authors and do not necessarily reflect the views of the John Templeton Foundation. We also acknowledge support from CoQuSy project PID2022-140506NB-C21, and the María de Maeztu project CEX2021-001164-M for Units of Excellence, funded by MICIU/AEI/10.13039/501100011033/FEDER, UE. GM acknowledges the Ram\'on y Cajal program RYC2021-031121-I funded by MICIU/AEI/10.13039/501100011033 and European Union NextGenerationEU/PRTR.  This research was also supported by the FWF (Austrian Science Fund) through an Erwin Schrödinger Fellowship (Project J 4699).

\bibliography{References}

\clearpage


\appendix

\section{General encoding for the transmitted information between labs} \label{appA}

\color{black}
In this Appendix we consider generic scenarios allowing that some information about the qubit measurement results in the friend's laboratory becomes accessible to Wigner. More precisely, we consider unitary interactions that partially correlate the accessible environmental states in the friend's laboratory $\{\ket{\phi_0}_F, \ket{\phi_1}_F \}$ (containing univocal information about the outcome of the qubit measurement) with those in Wigner's laboratory $\{\ket{\omega_0}_W, \ket{\omega_1}_W \}$. This can be achieved by generalizing the unitary that couples the system, the friend's lab and Wigner's environment defined through Eq.~\eqref{collmeas1} by allowing it to depend on a parameter $\theta$, i.e. $U_\mathrm{int}(\theta)$ with $\theta \in [0,1]$, so that its relevant action over the states of the system, the friend and Wigner's environments is now:
\begin{equation}
\begin{aligned} \label{collmeas3}
     {U_{\rm int}}(\theta) &  \ket{\uparrow}_S\ket{\phi_0}_F\ket{\omega_0}_W     \\ &{=}~  \frac{1}{\sqrt{2}} ( \ket{\Phi_\theta^{+}}_{SF} \ket{\omega_0}_W + \ket{\Phi_\theta^{-}}_{SF} \ket{\omega_1}_W ), \\ 
     {U_{\rm int}}(\theta) &\ket{\downarrow}_S\ket{\phi_1}_F\ket{\omega_0}_W \\ 
     &{=}~    \frac{1}{\sqrt{2}} ( \ket{\varphi_\theta^{+}}_{SF} \ket{\omega_0}_W - \ket{\varphi_\theta^{-}}_{SF} \ket{\omega_1}_W ).
\end{aligned}
\end{equation}
Here we introduced the following (unnormalized) parameterized states of the system and the friend's lab: 
\begin{align}
\ket{\Phi_\theta^\pm}_{SF} &= \frac{1}{\sqrt{2}} \Big( [1 \pm \sin\left( \frac{\pi \theta}{2}\right)] \ket{\uparrow}_S\ket{\phi_0}_F \nonumber \\ & ~~~~\pm \cos \left( \frac{\pi \theta}{2}\right) \ket{\downarrow}_S\ket{\phi_1}_F \Big), \\
\ket{\varphi_\theta^\pm}_{SF} &= \frac{1}{\sqrt{2}} \Big( \cos \left( \frac{\pi \theta}{2}\right)\ket{\uparrow}_S\ket{\phi_0}_F \nonumber \\ & ~~~~\pm   [1 \mp \sin\left( \frac{\pi \theta}{2}\right)] \ket{\downarrow}_S\ket{\phi_1}_F \Big),
\end{align}
{with $\langle \Phi^{\pm}|\Phi^{\pm}\rangle= \langle \varphi^{\mp}|\varphi^{\mp}\rangle = [1 \pm \sin(\pi\theta/2)]$}, which reduce to the two standard Bell states given below Eq.~(\ref{collmeas1}) for $\theta = 0$. Henceforth it is easy to check that the above unitary $U_\mathrm{int}(\theta)$ reduces to Eq.~\eqref{collmeas1} for the case $\theta = 0$. On the other hand, for $\theta > 0$
such a unitary generically encodes partial information about the qubit measurement results in friend's lab into superpositions of the  states of Wigner's environment with different amplitudes that depend on $\theta$. In the extreme case $\theta=1$ we have $\ket{\Phi_1^+}_W = \ket{\uparrow_0}_S \ket{\phi_0}_F$ and $\ket{\varphi_1^-}_W = \ket{\downarrow_0}_S \ket{\phi_1}_F$ (while $\ket{\Phi_1^-} = \ket{\varphi_1^+} =0$) so that Wigner's accessible lab states become perfectly correlated with the measured qubit states and friend's lab results.  

Applying $U_\mathrm{int}(\theta)$ to Wigner's state in Eq.~\eqref{psi1} we obtain the unnormalized (c.f. Eq.~\eqref{psiU2}):
\begin{align} \label{W2app}
    \ket{\Psi_2^W} &= \frac{1}{2} \big[ (\ket{\Phi_\theta^+}_{SF} + \ket{\varphi_\theta^+}_{SF})\ket{\omega_0} \nonumber \\ 
     &~~~~~+ (\ket{\Phi_\theta^-}_{SF} - \ket{\varphi_\theta^-}_{SF}) \ket{\omega_1}  \big]. 
\end{align}   
Using the explicit form of the states $\ket{\Phi_\theta^\pm}$ and $\ket{\varphi_\theta^\pm}$ that state can be rewritten as
 \begin{align}   
    \ket{\Psi_2^W}&=\frac{1}{2\sqrt{2}}\Big(\big[ 1 + \sin\left( \frac{\pi \theta}{2}\right) + \cos \left( \frac{\pi \theta}{2}\right) \big] \ket{\uparrow}_S\ket{\phi_0}_F \ket{\omega_0}_W \nonumber \\ 
    &+  \big[ 1 - \sin \left( \frac{\pi \theta}{2}\right) + \cos \left( \frac{\pi \theta}{2}\right) \big] \ket{\downarrow}_S\ket{\phi_1}_F \ket{\omega_0}_W \nonumber \\ 
   &+  \big[ 1 - \sin \left( \frac{\pi \theta}{2}\right) - \cos \left( \frac{\pi \theta}{2}\right) \big] \ket{\uparrow}_S\ket{\phi_0}_F \ket{\omega_1}_W \nonumber \\ 
   &+  \big[ 1 + \sin \left( \frac{\pi \theta}{2}\right) - \cos \left( \frac{\pi \theta}{2}\right) \big] \ket{\downarrow}_S\ket{\phi_1}_F \ket{\omega_1}_W \Big).
\end{align}
Therefore, depending on the value of $\theta$, Wigner's environmental states $\ket{\omega_0}_E$ and $\ket{\omega_1}_E$ can respectively get correlated to the system and the friend's lab states $\ket{\uparrow}_S\ket{\phi_0}_F$ and $\ket{\downarrow}_S\ket{\phi_1}_F$. In particular, the correlation is stronger, the larger the value of $\theta$ is, while for $\theta = 0$ the two last lines of the state above disappear and we recover Eq.~\eqref{psiU2}. 

In the case $\theta=1$ the state in Eq.~\eqref{W2app} reduces to: 
\begin{equation} \label{W2app2}
{\ket{\Psi_2^W}}= \frac{1}{\sqrt{2}}\Big(\ket{\uparrow}_S \ket{\phi_0}_F \ket{\omega_0}_W + \ket{\downarrow}_S \ket{\phi_1}_F \ket{\omega_1}_W \Big)    
\end{equation}
corresponding to perfect correlations between the measurement results and both friend's and Wigner's environmental states. In this case by looking at the preferred basis of Wigner's environment we obtain with $1/2$ probabilities the states $\ket{\uparrow}_S \ket{\phi_0}_F \ket{\omega_0}_W$ and $\ket{\downarrow}_S \ket{\phi_1}_F \ket{\omega_1}_W$.

On the other hand, applying $U_\mathrm{int}(\theta)$ to the friend's (collapse-rule) state assignment in Eq.~\eqref{measm} gives us the two following states:
\begin{equation}
    \label{F2app}
  \ket{\Psi_2^F}= 
\begin{cases}
    \frac{1}{\sqrt{2}} ( \ket{\Phi_\theta^{+}}_{SF} \ket{\omega_0}_W + \ket{\Phi_\theta^{-}}_{SF} \ket{\omega_1}_W ), \\ \vspace{0.1cm} \\
     \frac{1}{\sqrt{2}} ( \ket{\varphi_\theta^{+}}_{SF} \ket{\omega_0}_W - \ket{\varphi_\theta^{-}}_{SF} \ket{\omega_1}_W ),
\end{cases}
\end{equation}
for outcomes $\phi_0$ and $\phi_1$, respectively, to be compared with the state $\ket{\Psi_2^F}$ in Eq.~\eqref{measU2}.
It should be noticed here that in this case the environmental states $\ket{\omega_0}$ and $\ket{\omega_1}$ are partially correlated with $\ket{\uparrow}_S\ket{\phi_0}_F$ and $\ket{\downarrow}_S\ket{\phi_1}_F$ as long as $\theta > 0$. If $\theta= 1$ we obtain:
\begin{equation}
    \label{F2app2}
  \ket{\Psi_2^F}= 
\begin{cases}
   \ket{\uparrow}_S\ket{\phi_0}_F \ket{\omega_0}_W ~~~{\rm for~ outcome}~ \phi_0, \\ \vspace{0.1cm} \\
    \ket{\downarrow}_S\ket{\phi_1}_F \ket{\omega_1}_W ~~~{\rm for~ outcome}~ \phi_1,
\end{cases}
\end{equation}
with perfect correlations between the preferred basis states of the friend's and Wigner lab degrees of freedom. Therefore the information about the friend's measurement outcomes is completely accessible from Wigner's environmental states.

\section{Wigner's and friend's predictions for partial leakage of information} \label{appB}

Let's now discuss the results of the checking procedure presented in Secs.~\ref{sec:measW} and \ref{sec:measF} for measurements in Wigner's and friend's bubbles, respectively, in the case of general encoding of information presented in Appendix~\ref{appA}. 

In the case of measurement $M_W$ in Wigner's preferred basis $\{\ket{\omega_0}, \ket{\omega_1} \}$, based on state assignment in Eq.~\eqref{W2app} the prediction of Wigner for the two outcomes of the measurement are:
\begin{equation}
    \label{W2appM}
  \ket{\Psi_3^W}= 
\begin{cases}
  \frac{1}{\sqrt{2}}(\ket{\Phi_\theta^+}_{SF} + \ket{\varphi_\theta^+}_{SF})\ket{\omega_0}_W ~~~{\rm for~ outcome}~ \omega_0, \\ \vspace{0.1cm} \\
   \frac{1}{\sqrt{2}} (\ket{\Phi_\theta^-}_{SF} - \ket{\varphi_\theta^-}_{SF})\ket{\omega_1}_W  ~~~{\rm for~ outcome}~ \omega_1,
\end{cases}
\end{equation}
with respective probabilities $p_{\omega_0}^W = [1+ \cos(\pi \theta/2)]/2$ and $p_{\omega_1}^W = [1 - \cos(\pi \theta/2)]/2$,  which are submitted to the referee $R_W$. Notice that now the prediction of Wigner is not deterministic, but biased in general towards $\omega_0$, as $1/2 \leq p_{\omega_0}^W \leq 1$ and $0 \leq p_{\omega_1}^W \leq 1/2$.
On the other hand, the friend's prediction of the probabilities for observing one or the other outcome of $M_W$, basing on state assignment Eq.~\eqref{F2app}, are again $p^F_{\omega_0}=1/2$ and $p^F_{\omega_1}=1/2$, which are also submitted to $R_W$. Notice that if no information about which outcome is present in the encoding, i.e. $\theta = 0$, this reduces to the case discussed in Sec.~\ref{sec:measW}. However, we observe that even if a part of the “which-outcome" information is leaked from the friend's to Wigner's environment, their predictions of the estimated probabilities of the two observers will still differ and hence the referee $R_W$ by performing the measurement $M_W$ can detect this mismatch performing a sufficient number of runs. Interestingly Wigner's and the friend's predictions become closer to each other as more information about the qubit measurement is leaked in the interaction between labs (which corresponds to increasing values of $\theta$) reaching the case of equal predictions for $\theta = 1$. 

\begin{figure}[t!]
    \centering
    \includegraphics[width=1.0\columnwidth]{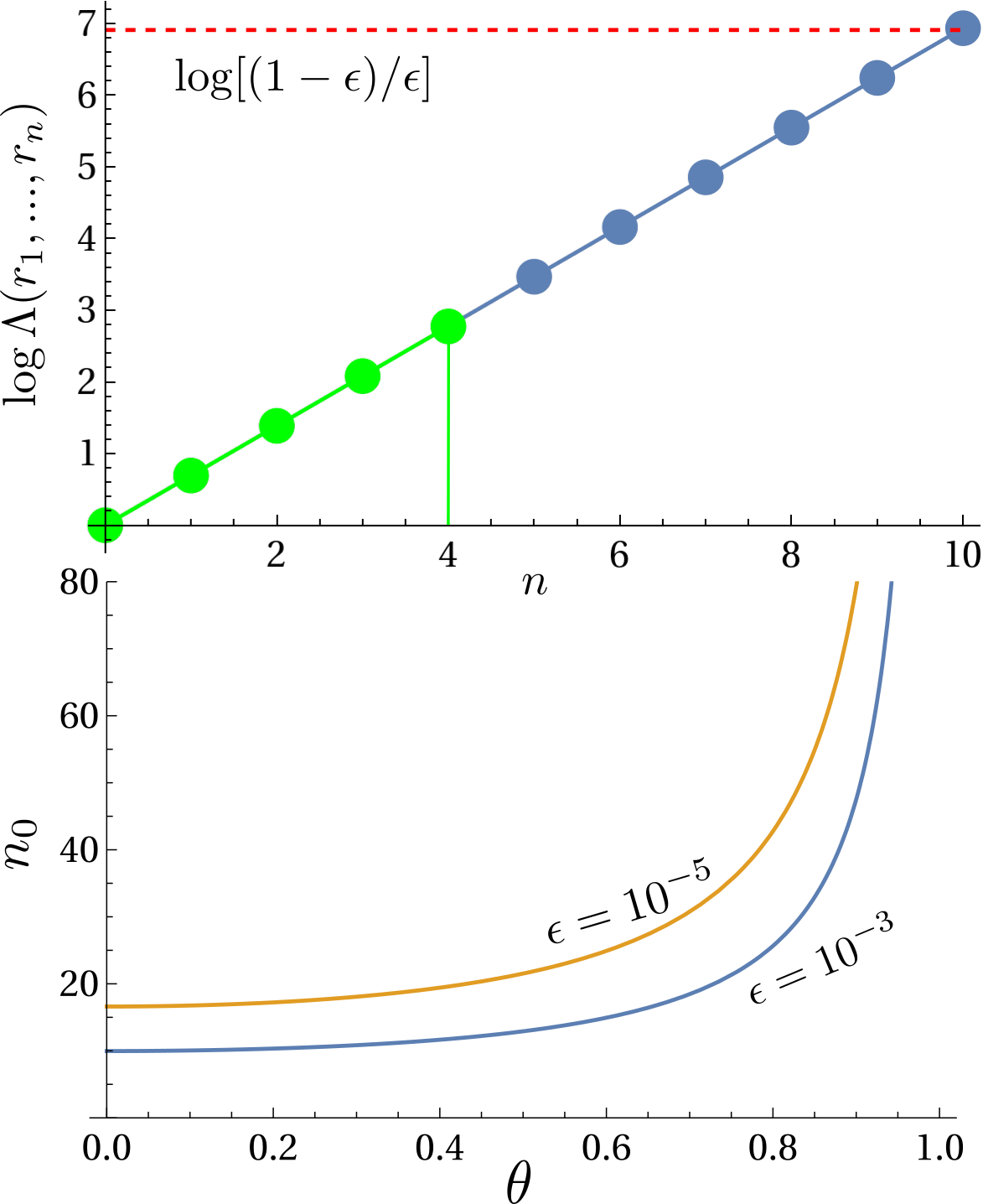}
    \caption{Top panel: Two examples of Wald's sequential probability ratio test for $\theta = 0$ (case in the main text). The blue dots represents the log-likelihood ratio for sequence of consecutive $\omega_0$ outcomes, reaching the acceptance threshold of Wigner's prediction (dashed line) for $\epsilon= 10^{-3}$ at run $n=10$. The green dots represent instead the log-likelihood ratio in the eventual case of a sequence of 4 consecutive $\omega_0$ values followed by $\omega_1$, leading to the acceptance of the friend's prediction at run $n=5$. Bottom panel: Minimum number of runs $n_0$ leading to outcomes $\omega_0$ in the measurement $M_W$ needed to validate Wigner's prediction as a function of the encoding parameter $\theta$ in the range $[0, 1]$ for two different values of the failure probability $\epsilon$.}
    \label{fig}
\end{figure}

In particular, we consider Wald's sequential probability ratio test~\cite{Wald1945}, one of the three standard approaches to hypothesis testing. There, two hypotheses for the probability of a sequence of observed outcomes $r$ (here $p_r^W$ and $p_r^F$ with $r=\{\omega_0, \omega_1\}$) are compared by constructing their log-likelihood ratio and computing its evolution as the number of observations increases. One of the two hypotheses is accepted if the log-likelihood reaches predefined positive or negative thresholds. In our situation, the referee $R_W$ may collect statistics from consecutive runs of the game and construct the cumulative log-likelihood ratio between Wigner's and friend predictions. After $n$ runs obtaining outcomes $\{ r_1, r_2, ..., r_n \}$ the cumulative log-likelihood ratio would read:
\begin{equation}
\begin{aligned}
    &\log \Lambda(r_1, ..., r_n) := \log \frac{p_{r_1}^W p_{r_2}^W ... p_{r_n}^W}{p_{r_1}^F p_{r_2}^F ... p_{r_n}^F} \\ &= n_0 \log [1+ \cos \left(\frac{\pi \theta}{2}\right)]+ n_1 \log [1- \cos \left(\frac{\pi \theta}{2}\right)],
\end{aligned}
 \end{equation}
where $p^W_{r_i}$ ($p^F_{r_i}$) are the probability assignments from Wigner (friend) perspective for outcome $r_i = \{ \omega_0, \omega_1\}$, and in the second line $n_0$ stands for the number of outcomes $\omega_0$ obtained and $n_1$ for the number of outcomes $\omega_1$. That quantity increases as outcomes $r_i = \omega_0$ in the sequence are accumulated in consecutive runs, and decreases when outcomes $r_i = \omega_1$ are detected (see Fig.~\ref{fig}). 
The referee will award Wigner as soon as $\log \Lambda$ first reaches the upper threshold $\log [(1-\epsilon)/\epsilon] > 0$, where $\epsilon > 0$ is the (typically small) error of test failure. On the contrary, the referee will award the friend as soon as $\log \Lambda$ reaches the lower threshold given by $\log [\epsilon/(1-\epsilon)] < 0$. For the case $\theta = 0$ the lower thresholds will be automatically reached as soon as the first outcome $r_i = \omega_1$ is detected since it is fully incompatible with Wigner's prediction. On the other hand, for $\theta=1$ where the predictions of Wigner and her friend become equal, $\log \Lambda = 0$ for any sequence and hence awarding one or the other is not possible.

The minimum 
number of runs leading to outcome $\omega_0$ needed to validate Wigner's prediction (occurring in the case when no outcome $\omega_1$ is obtained) would be:
\begin{equation}
    n_0 = \frac{\log [(1 - \epsilon)/\epsilon]}{ \log[1+ \cos \left(\frac{\pi \theta}{2}\right)]},
\end{equation}
which increases monotonically with $\theta$ (see Fig.\ref{fig}). In the limit $\theta \rightarrow 1$ where the “which-outcome" information is unambiguously leaked to Wigner's environment, the number of necessary runs diverges and both descriptions (unitary evolution and state collapse) perfectly agree. 

A similar situation is obtained in the case of performing the measurement $M_F$ in the friend's bubble. Similarly to the situation described in Sec.~\ref{sec:measF} when a second measurement of the qubit state is performed in the friend's lab the results will confirm the initial measurement state assignment: she predicts $p^F_{\phi_0}=1$ (or 0), $p^F_{\phi_1}=0$ (or 1) and gives this prediction to referee R$_{F}$. On the other hand, from the point of view of Wigner, the state after the repeated measurement using unitary $U_{\rm meas}$ between system and an extra degree of freedom of the friend's lab $F^\prime$ reads:
\begin{align}
    & \ket{\Psi^W_4}  = (\hat{U}_{\mathrm{meas}} \otimes \mathds{1}_{F}) \ket{\Psi^{W}_2} \ket{\phi_0}_{F^\prime} \\
 & =\frac{1}{2\sqrt{2}}\Big(\big[ 1 + \sin\left( \frac{\pi \theta}{2}\right) + \cos \left( \frac{\pi \theta}{2}\right) \big] \ket{\uparrow}_S\ket{\phi_0}_F \ket{\phi_0}_{F'} \ket{\omega_0}_W \nonumber \\ 
    &+  \big[ 1 - \sin \left( \frac{\pi \theta}{2}\right) + \cos \left( \frac{\pi \theta}{2}\right) \big] \ket{\downarrow}_S\ket{\phi_1}_F \ket{\phi_1}_{F'} \ket{\omega_0}_W \nonumber \\ 
   &+  \big[ 1 - \sin \left( \frac{\pi \theta}{2}\right) - \cos \left( \frac{\pi \theta}{2}\right) \big] \ket{\uparrow}_S\ket{\phi_0}_F \ket{\phi_0}_{F'} \ket{\omega_1}_W \nonumber \\ 
   &+  \big[ 1 + \sin \left( \frac{\pi \theta}{2}\right) - \cos \left( \frac{\pi \theta}{2}\right) \big] \ket{\downarrow}_S\ket{\phi_1}_F \ket{\phi_1}_{F'} \ket{\omega_1}_W \Big). \nonumber
\end{align}
The probabilities for the outcomes $\uparrow$ and $\downarrow$ from Wigner's perspective are $p^W_{\phi_0}=(1 + \sin (\pi \theta/2)\cos(\pi \theta/2))/2$ and $p^W_{\phi_1}=(1 - \sin (\pi \theta/2)\cos(\pi \theta/2))/2$ from the above state, which are sent to referee R$_F$ inside the friend's bubble. As we can see in this situation the Referee $R_F$ can also see a difference between Wigner and friend's predictions for the measurement $M_F$ whenever $\theta < 1$. Again, as expected, when perfect “which-outcome" information is available in Wigner's environment $(\theta = 1)$ both descriptions becomes equivalent.

\end{document}